# Non-Invasive Fault Detection of Stator Windings of Induction Motors

Fairuz, Rayyan Bin

## Abstract

Condition monitoring of induction motor has been widely researched over recent years due to its ability to monitor operating characteristics and the health status of induction motor. Various methods have been used to monitor induction motors such as thermal monitoring and vibration analysis. This paper introduces an alternative method which is to use an inductive coupling method to extract in-circuit impedance of induction motor. This method allows for an online measurement of the system under test (SUT) preventing any unnecessary shutdowns. These unnecessary shutdowns may incur a loss of revenue for relevant industries that depends heavily on induction motor operations. Two sets of experiments were conducted in this paper, the first experiment was to examine the accuracy of the proposed method by simulating various SUT using simulated resistor, inductor, and capacitor (RLC) network. The next experiment was to detect incipient stator faults such as turn to turn faults in an induction motor. This proposed method measures the impedance of the stator winding of an induction motor to identify abnormalities. In addition, this method allows for an accurate in-circuit impedance measurement without the influence of motor load and frequency changes as well as faults such as bearing and rotor faults.

## I. Introduction

The induction motor has been widely used machine in various industrial applications such as propulsion motor in electric cars, electric-powered trains, compressors and many more [1]. These motors have high reliability, simple construction and are low cost. In most situations, the induction motor has to operate continuously for a long period. Continuous usage of these machines will eventually cause faults to occur. Faults such as stator fault, bearing fault and cracked rotor bar are the most common faults that occur in induction motors [2].

These faults are detected either when the motors are due for periodic maintenance or when the motor has already failed. Such unexpected failures of the induction motor can cause great economic losses. Hence, condition monitoring of the induction motor is necessary to detect earlier signs of failures [3]. Non-invasive condition monitoring techniques are preferred, as they will not disturb the operation of the entire system. This paper introduces an advanced technique of condition monitoring of induction motor to detect early signs of stator fault. The proposed technique is based on the inductive coupling method [4-6]. Stator fault occurs when insulation between windings has failed which then causes a short circuit in the system. This results in an unbalance characteristic in the overall impedance of the induction motor.

This paper is organized as follows: Section II describes the inductive coupling method with theories behind. Section III describe the experiments conducted to validate the inductive coupling method for fault detection of induction motors. Feasibility and robustness test of this method were also carried out. Section IV concludes this paper.

## II. Methodology

This paper introduces an advanced technique of evaluating the operating characteristic and health condition of an induction motor by measuring the in-circuit impedance of the system. The in-circuit impedance of the induction motor will provide a base impedance value of a healthy operating machine. This base value can then be compared to an impedance value of an induction motor that has suffered stator faults. To measure the impedance of these machines, an inductive coupling method was used. The inductive coupling method compromises two clamped-on type current probes, an injecting inductive probe (IIP) and a receiving inductive probe (RIP) [7]. These probes are essentially inductive coupled transformers which induce a voltage across the wires of the system under test (SUT) through electromagnetic induction. The in-circuit impedance of the SUT can be extracted from the feedback signal from the RIP with the use of the two-port ABCD network approach [8]. This technique allows for a non-invasive impedance measurement as it does not require direct contact with the SUT removing the possibility of electrical hazard [9]. The clamped-on type inductive probes can also be easily installed around cables of the SUT for online impedance measurement.

Figure 1 illustrates the measurement set up using the inductive coupling measurement method. Components used were a 2 pole, 220-240V, 50Hz induction motor (IM), a computer-controlled signal generation and acquisition system (SGAS), clamped type receiving inductive probe (RIP), injecting inductive probe (IIP) and LabVIEW software. The induction motor is powered by an AC source (Vs) through wire connections. The wire passes through the RIP and IIP and connects to a single impedance represented by ($Z_{total}$). $Z_{total}$ compromises of impedance from induction motor, the internal impedance of voltage source and the impedance of the wires. The SGAS consists of a signal generation card (SGC) which outputs a sinusoidal signal voltage ($V_{sig}$) and frequency ($f_{sig}$) in through the IIP. The IIP will induce a small current through the circuit which will then be detected by the RIP. The result of the detected signal will be used to calculate the total impedance ($Z_{total}$) in the circuit. The feedback of the IIP and RIP is monitored by channel 1 (C1) and 2 (C2) of the signal acquisition card (SAC) of the SGAS through oscilloscope form the LabVIEW software.

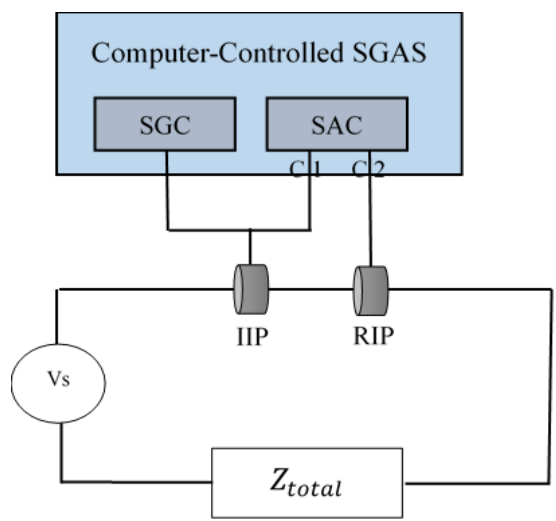

Figure 1. Basic measurement setup for impedance measurement of SUT.

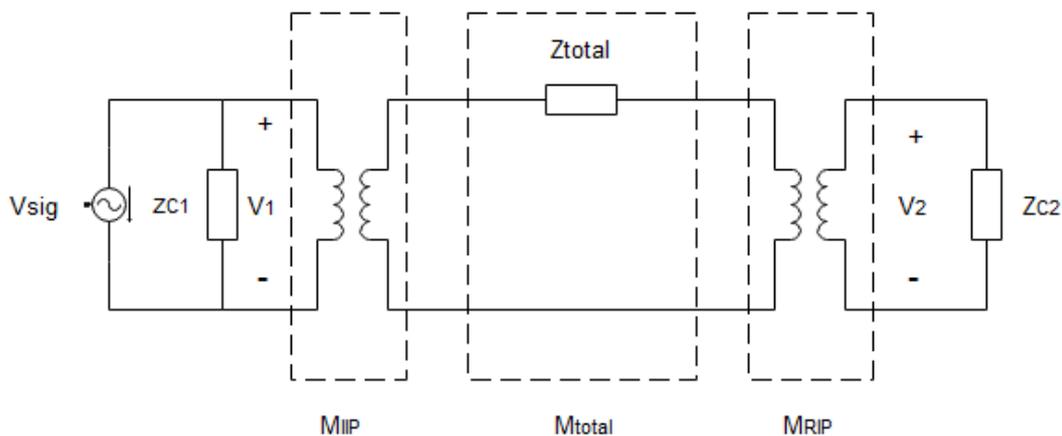

Figure 1. Equivalent circuit impedance of Figure 1 represented in two-port network.

The measurement in Figure 1 can be represented in a two-port cascaded ABCD network as shown in Figure 2. The output impedance of C1 and C2 is represented by $Z_{C1}$ and $Z_{C2}$ which has an impedance value of 1 MΩ and 50 Ω respectively. $M_{IIP}$, $M_{total}$, and $M_{RIP}$ represent the ABCD network of the IIP, $Z_{total}$ and RIP respectively. Thus, the input to output relation can be express by:

$$\begin{bmatrix} V1 \\ I1 \end{bmatrix} = \begin{bmatrix} A_{IIP} & B_{IIP} \\ C_{IIP} & D_{IIP} \end{bmatrix} \begin{bmatrix} A_{total} & B_{total} \\ C_{total} & D_{total} \end{bmatrix} \begin{bmatrix} A_{RIP} & B_{RIP} \\ C_{RIP} & D_{RIP} \end{bmatrix} \begin{bmatrix} V2 \\ I2 \end{bmatrix} \quad (1)$$

Given that $I_2 = \dfrac{V2}{50}$ and $M_{total}$ can be express as [10]:

$$M_{total} = \begin{bmatrix} 1 & Z_{total} \\ 0 & 1 \end{bmatrix} \quad (2)$$

From the above two equations, $Z_{total}$ can be determined by:

$$Z_{total} = \dfrac{1}{A_{IIP}\,(C_{RIP}+D_{RIP}/50)} \cdot \dfrac{V1}{V2} - \dfrac{A_{RIP}+B_{RIP}/50}{C_{RIP}+D_{RIP}/50} - \dfrac{B_{IIP}}{A_{IIP}} \quad (3)$$

$V_1$ and $V_2$ can be determined directly from the setup. The ABDC parameters of the IIP and RIP have to be pre-characterised using a vector network analyser (VNA) with a test jig [8]. Once the parameters have been set, $Z_{total}$ can be found from equation (3).

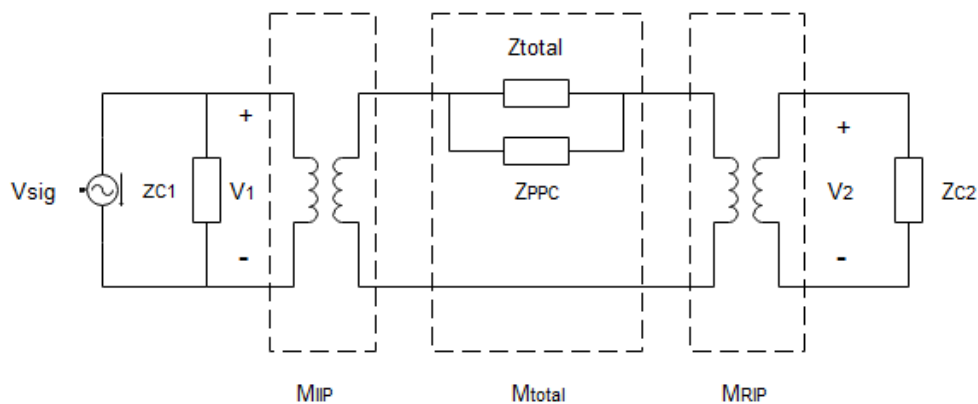

Figure 2. Equivalent circuit impedance of Figure 1 with probe to probe coupling.

The use of IIP and RIP causes probe to probe coupling (PPC) inaccuracy [11]. The distance (*d*) between the IIP and RIP causes undesired impedance. As a result, the IIP and RIP have to be calibrated to eliminate potential measurement errors. A calibration jig is required for the calibration of the IIP and RIP using open short and load calibration (OSL). Figure 3 shows the effect of undesired impedance resulting from PPC effect represented by $Z_{PPC}$. With the equation in (3) and considering $Z_{PPC}$:

$$Z_{total} // Z_{PPC} = \frac{1}{A_{IIP}\,(C_{RIP}+D_{RIP}/50)} \cdot \frac{V1}{V2} - \frac{A_{RIP}+B_{RIP}/50}{C_{RIP}+D_{RIP}/50} - \frac{B_{IIP}}{A_{IIP}} \qquad (4)$$

Thus, the measurement error $\Delta Z_{total}$ due to $Z_{PPC}$ can be calculated by:

$$\Delta Z_{total} = \frac{Z_{total}Z_{PPC}/(Z_{total}+Z_{PPC})-Z_{total}}{Z_{total}} \qquad (5)$$

Equation (5) can be simplified to:

$$\Delta Z_{total} = \frac{-Z_{total}}{Z_{total}+Z_{PPC}} \qquad (6)$$

From (6), $\Delta Z_{total} \approx 0$ when $Z_{total} \ll Z_{PPC}$. Therefore the measurement error is negligible. However, when $Z_{PPC}$ is significant, the error is non-negligible. An OSL calibration method is used to eliminate this error. Equation (4) is simplified to:

$$Z_{total}\,Z_{PPC}\,/\,(Z_{total}+Z_{PPC}) = k \cdot \frac{V1}{V2} + b \qquad (7)$$

where

$$k = \frac{1}{A_{IIP}(C_{RIP}+D_{RIP}/50)} \qquad (8)$$

$$b = -\frac{A_{RIP}+B_{RIP}/50}{C_{RIP}+D_{RIP}/50} - \frac{B_{IIP}}{A_{IIP}} \qquad (9)$$

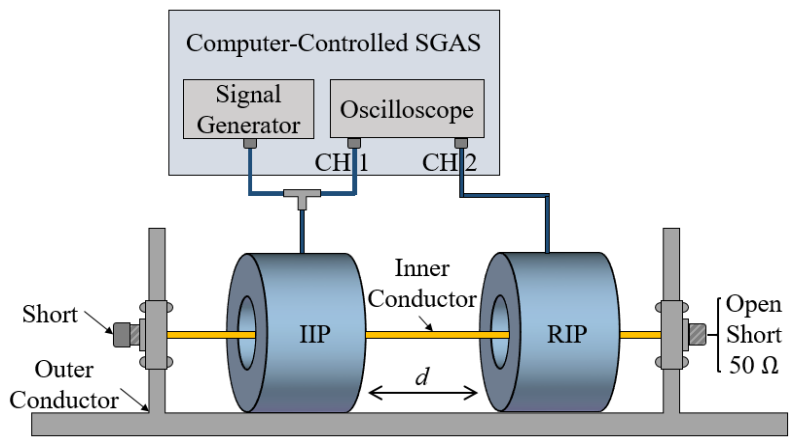

Figure 3. OSL calibration fixture.

An OSL calibration jig is required for the calibration of the IIP and RIP as shown in Figure 4. The OSL calibration jig is used to determine the values of k, b, and $Z_{PPC}$. One end of the fixture is left shorted and the other end (calibration end) is used for the calibration of open, short and load at 50 Ω. A test signal with frequency ($f_{sig}$) which is produced by the SGAS is injected through the IIP and the feedback is monitored by the RIP. When the calibration end of the fixture is left open $Z_{total} = \infty$, then (7) will be:

$$Z_{PPC} = k \cdot \frac{V1}{V2}|_{opened} + b \tag{10}$$

When the calibration end of the fixture is left open $Z_{total} = 0$, then (7) will be:

$$0 = k \cdot \frac{V1}{V2}|_{shorted} + b \tag{11}$$

When the calibration end of the fixture is left open, $Z_{total}$ = 50 Ω, then (7) will be:

$$50 \cdot Z_{PPC} / (50 + Z_{PPC}) = k \cdot \frac{V1}{V2}|_{50} + b \tag{12}$$

Thus k, b, $Z_{PPC}$ can be determined by the following expressions:

$$k = \frac{50\left(\frac{V1}{V2}|_{opened} - \frac{V1}{V2}|_{50}\right)}{\left(\frac{V1}{V2}|_{opened} - \frac{V1}{V2}|_{shorted}\right)\left(\frac{V1}{V2}|_{50} - \frac{V1}{V2}|_{shorted}\right)} \tag{13}$$

$$b = \frac{50 \cdot \frac{V1}{V2}|_{shorted}\left(\frac{V1}{V2}|_{50} - \frac{V1}{V2}|_{opened}\right)}{\left(\frac{V1}{V2}|_{opened} - \frac{V1}{V2}|_{shorted}\right)\left(\frac{V1}{V2}|_{50} - \frac{V1}{V2}|_{shorted}\right)} \tag{14}$$

$$Z_{PPC} = \frac{50\left(\frac{V1}{V2}|_{opened} - \frac{V1}{V2}|_{50}\right)}{\left(\frac{V1}{V2}|_{50} - \frac{V1}{V2}|_{shorted}\right)} \tag{15}$$

Substituting (13)-(15), $Z_{total}$ can be expressed by:

$$Z_{total} = 50 \cdot \frac{\left(\frac{V1}{V2}|_{opened} - \frac{V1}{V2}|_{50}\right)\left(\frac{V1}{V2} - \frac{V1}{V2}|_{shorted}\right)}{\left(\frac{V1}{V2}|_{50} - \frac{V1}{V2}|_{shorted}\right)\left(\frac{V1}{V2}|_{opened} - \frac{V1}{V2}\right)} \tag{16}$$

Equation (16) is only able to use if voltages $V_1$ and $V_2$ are measurable. $V_2$ depends on $Z_{total}//Z_{PPC}$. When $Z_{total}$ = 0 or 50 Ω, $Z_{total} << Z_{PPC}$. Therefore $Z_{total}// Z_{PPC} \approx Z_{total}$. When $Z_{total} = \infty$, $V_2|_{opened}$ depends on $Z_{PPC}$. If $Z_{PPC}$ is finite then $V_2|_{opened}$ is measurable. If $Z_{PPC}$ goes to infinity, $V_2|_{opened}$ will be unmeasurable due to the background harmonics in the oscilloscope of the SGAS. Thus the effect of PPC is negligible and (7) can be expressed as:

$$Z_{total} = k \cdot \frac{V_1}{V_2} + b \tag{17}$$

With the same calibration at short and 50 Ω, $Z_{total}$ can be finally expressed as:

$$Z_{total} = \frac{50 \cdot (\frac{V_1}{V_2} - \frac{V_1}{V_2}|_{shorted})}{(\frac{V_1}{V_2}|_{50} - \frac{V_1}{V_2}|_{shorted})} \tag{18}$$

Components such as power amplifier, surge protector and attenuator can be added in the measurement set up when experiments involve high voltage such as an induction motor. The overall impedance however, will not be affected and the impedance calculation formula mentioned above will be the same.

## III. Experimental Validation

Figure 5 shows the measurement setup of this experiment. A few additional components were required for this experiment, a power amplifier (PA), to amplify the input signal from the SGAS, two attenuators (AT) to reduce the amplitude level of an incoming signal and two surge protector (SP1 and SP2) to protect the SGAS from a sudden increase in voltage from the SUT. The SUT compromises of a three-phase induction motor (IM) connected to a variable frequency drive (VFD).

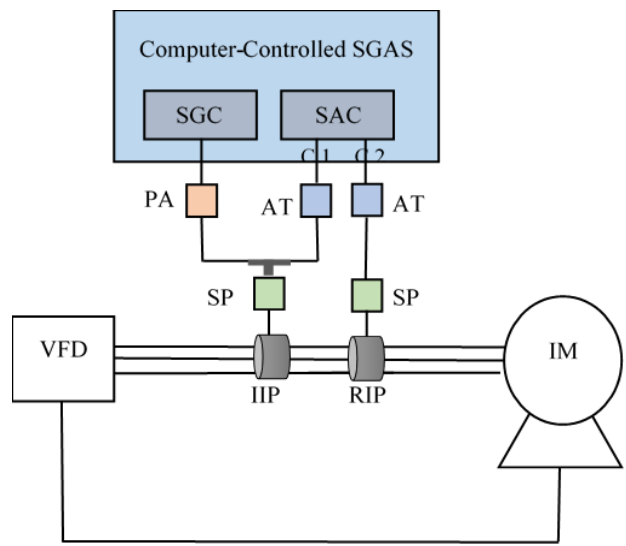

Figure 5. Proposed measurement setup.

Figure 6 shows the actual measurement setup for the experiment. The experiment is separated into four various measurements. The first measurement is to examine the effect of the impedance of the induction motor with respect to the rotational speed (Impedance Vs rotational

Speed). The second part was to examine the effect of impedance of the motor with respect to the load (Impedance Vs loading). The third part of the experiment was to examine the effect of rotor fault and bearing fault to the impedance of the motor (Impedance Vs bearing and rotor fault). Lastly was to induce a stator turn to turn fault and examine the effect on the impedance of the IM (Impedance Vs stator fault).

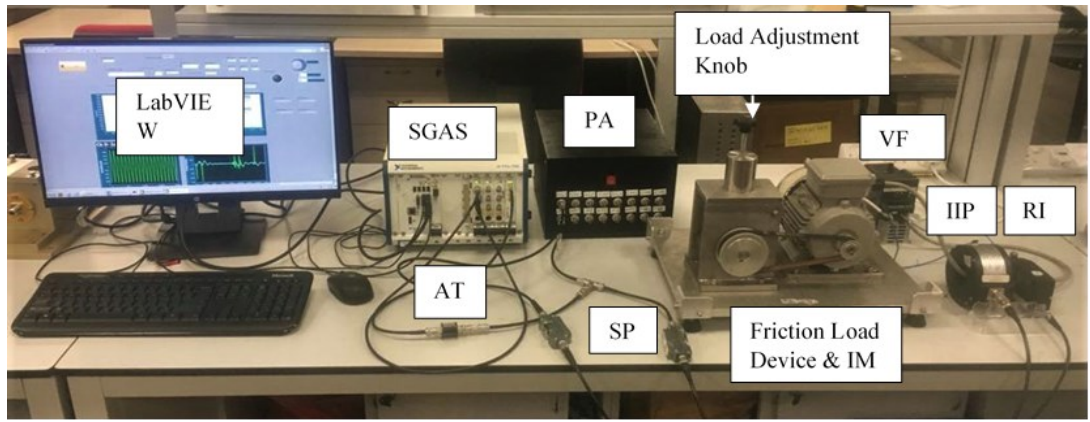

Figure 6. Actual measurement setup.

This experiment was conducted to examine the effect of the impedance of the motor when the rotational speed is increased. An impedance of a healthy induction motor at 0 RPM was measured as reference to be 6750Ω∠-67.9°. In this experiment, based on the current harmonics, an $f_{sig}$ of 91.3 kHz was chosen. This frequency contributes to low a noise level allowing optimal signal to noise ratio. A low signal voltage was also used this the following experiments. Table 1 shows the impedance of the induction motor with varying RPM from 1159 to 2943. The impedance measured at different RPM has an insignificant change to the impedance of a healthy induction motor at 0 RPM as shown by its relative change. The relative change is calculated as follows: Relative change $= \frac{|Z_{Healthy} - Z_{SUT}|}{|Z_{Healthy}|} \cdot 100\%$

Table 1. Impedance vs rotational speed.

| Impedance of induction motor $Z_{SUT}$ | | Rotational speed of induction motor (RPM) | Frequency of VFD (Hz) | Relative change (%) |
|---|---|---|---|---|
| $|Z|(\Omega)$ | $\angle Z(°)$ | | | |
| 6748.51 | -67.54 | 1159 | 20 | 0.022 |
| 6748.6 | -67.86 | 1755 | 30 | 0.021 |
| 6758.38 | -68.30 | 2346 | 40 | 0.118 |
| 6750.22 | -68.50 | 2943 | 50 | 0.003 |

This experiment was conducted to examine the effect of the impendence of the induction motor with varying loads. A friction load device was used to induce a loading effect to the induction motor as shown in Figure 7. A knob on the friction load device allows for a varying load on the induction motor. Three measurements were taken, measurement with no load, half load, and full load. The induction motor was set to a rotational speed of 2943 rpm with no load initially. Then to emulate half load and full load, the load adjustment know was rotated an approximate 4 revolutions and 8 revolutions respectively. As shown in Table 2, the impedance measured does not have significant change over the impedance measured for a healthy motor.

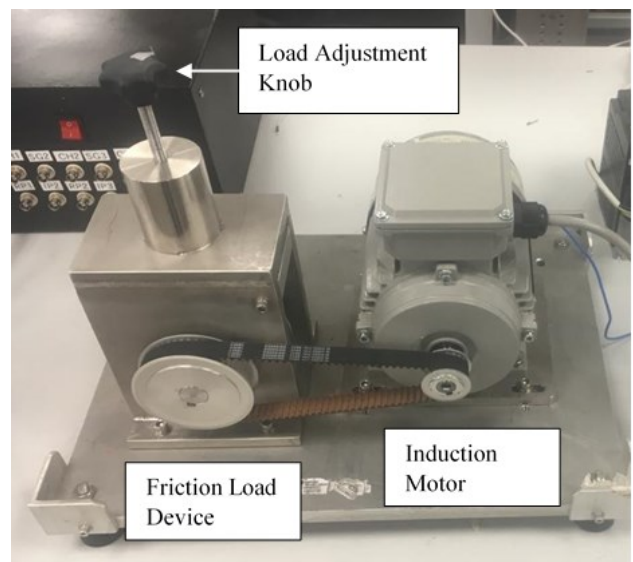

Figure 7. Friction load device.

Table 2. Impedance vs loading.

| Impedance of induction motor $Z_{SUT}$ | | Rotational speed of induction motor (RPM) | Friction Load | Relative change (%) |
|---|---|---|---|---|
| $|Z|(\Omega)$ | $\angle Z(°)$ | | | |
| 6750.22 | -68.50 | 2943 | No load | 0.003 |
| 6706.3 | -67.90 | 1946 | Half load | 0.652 |
| 6720.09 | -67.74 | 1142 | Full Load | 0.443 |

This experiment was conducted to see the effect of impedance of the inductor motor when it has suffered a broken rotor bar and a broken bearing. A hole was drilled at the centre of the rotor to emulate such rotor fault as shown in Figure 8. The result in Table 3 shows that there is no significant change in impedance of the induction motor to that of a healthy one.

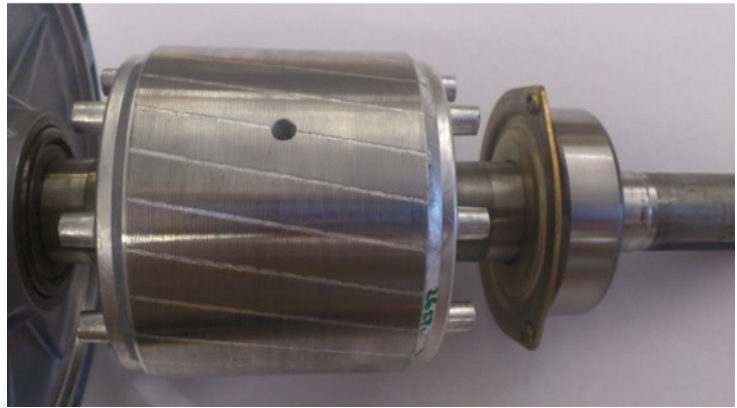

Figure 8. Induced rotor and bearing fault.

Table 3. Impedance vs bearing and rotor fault.

| Impedance of induction motor $Z_{SUT}$ | | Rotational speed of induction motor (RPM) | Frequency of VFD (Hz) | Relative change (%) |
| --- | --- | --- | --- | --- |
| $|Z|(\Omega)$ | $\angle Z(°)$ | | | |
| 6833.72 | -65.747 | 1145 | 20 | 1.24 |

The last experiment was to induce [Broken rotor] to turn fault buy grinding down the insulation coating on the copper wire of the stator. Figure 9 shows the copper wires that have been grinded down, soldered together, and secured with electrical tape. Once the fault has been induced, the induction motor was put back together and the impedance was measured. Table 4 shows the measured result of the induction motor that has suffered stator turn to turn fault. The measured value produces a significant change over the reference value at 4.23% deviation.

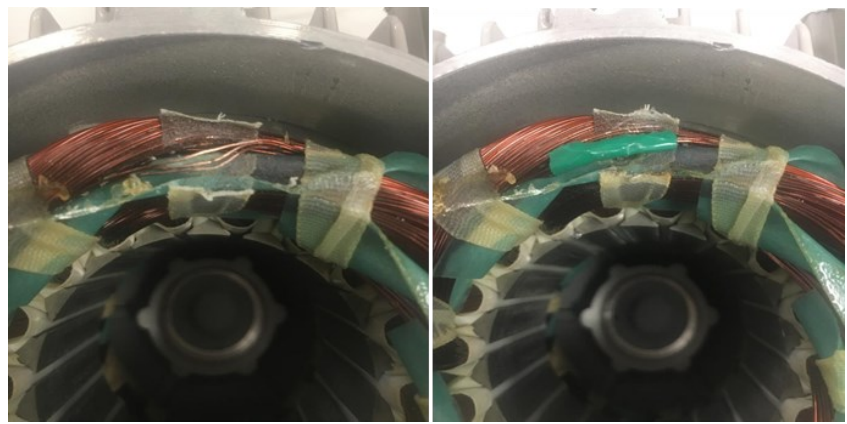

Figure 9. Induced stator turn to turn fault.

Table 4. Impedance vs stator fault.

| Impedance of induction motor $Z_{SUT}$ | | Rotational speed of induction motor (RPM) | Frequency of VFD (Hz) | Relative change (%) |
|---|---|---|---|---|
| $|Z|(\Omega)$ | $\angle Z(°)$ | | | |
| 7036 | -64.606 | 1155 | 20 | 4.237 |

## IV. Conclusion

This paper introduces the inductive coupling method for condition monitoring of the induction motor by measuring its in-circuit impedance. The inductive coupling method was validated with two main experiments. The first experiment was conducted with four different RLC networks with various *fsig* and distance *d*. The subsequent experiment was conducted with the inductive coupling method on an induction motor to diagnose stator turn to turn fault.

The inductive coupling method is a preferred method for condition monitoring of induction motor as it allows for online measurement. This means that this non-intrusive nature of this method would be favourable for industries that depends on the continuous operation of induction motor for revenue. This method has its advantage of being robust towards load change, speed variation, and bearing and rotor faults. This allows a more specific method of determining the condition of induction motors based on stator turn to turn faults.

In addition to its robustness, the inductive coupling method allows for easy installation to the cables of an induction motor with clamped-on type inductive probes. This allows for a more efficient preventive maintenance procedure for industry applications. The inductive coupling method discussed in This paper has can accurately measure the impedance of the stator winding of the induction motor to determine insipient stator turn to turn fault. Identifying insipient stator turn to turn fault, helps to prevent further damages to the induction motor which could result in its destruction.